\documentclass[debug,preprint,overfull]{epl}
\usepackage{amsmath,citesort}
\renewcommand{\thefootnote}{\fnsymbol{footnote}}

\newcommand{\fR}{\mathcal{R}}

\title{Correlation length of the 1D Hubbard Model at 
half-filling : equal-time one-particle Green's function}
\shorttitle{Correlation length of 1D Hubbard Model}
\author{Y.Umeno\inst{1}, M. Shiroishi\inst{2} \and A. Kl\"umper\inst{3}}
\institute{
  \inst{1} Department of Physics, 
           Graduate School of Science, University of Tokyo,\\
           Hongo 7-3-1, Bunkyo-ku, Tokyo 113, Japan \\
  \inst{2} Institute for Solid State Physics, University of Tokyo,
           Kashiwanoha 5-1-5, \\ Kashiwa, Chiba, 277-8581 Japan \\
  \inst{3} Theoretische Physik I, Universit\"{a}t Dortmund, 
Otto-Hahn-Str.~4, D-44221 Dortmund, Germany
}
\pacs{71.10.Fd}{Lattice fermion models (Hubbard model, etc.)}
\pacs{71.27.+a}{Strongly correlated electron systems; heavy fermions}
\pacs{05.30.Fk}{Fermion systems and electron gas}

\begin{document}

\maketitle

\begin{abstract}
  The asymptotics of the equal-time one-particle Green's function for
  the half-filled one-dimensional Hubbard model is studied at finite
  temperature. 
  We calculate its correlation length by evaluating the largest and 
  the second largest eigenvalues of the Quantum Transfer Matrix 
  (QTM). In order to allow for the genuinely fermionic nature of the 
  one-particle Green's function, we employ the fermionic formulation 
  of the QTM based on the fermionic ${R}$-operator of the Hubbard 
  model. The purely imaginary value of the second largest eigenvalue 
  reflects the ${k_{\rm F} (= \pi/2)}$ oscillations of the 
  one-particle Green's function at half-filling.  By solving 
  numerically the Bethe Ansatz equations with Trotter numbers up to 
  ${N=10240}$, we obtain accurate data for the correlation length at 
  finite temperatures down into the 
  very low temperature region. The 
  correlation length remains finite even at ${T=0}$ due to the 
  existence of the charge gap. Our numerical data confirm Stafford 
  and Millis' conjecture regarding an analytic expression for the 
  correlation length at ${T=0}$.
\end{abstract}

The one-dimensional (1D) Hubbard model 
\begin{equation}
\mathcal{H} = - \sum_{j=1}^L \sum_{\sigma = \uparrow \downarrow}     
       (c_{j+1 \sigma}^{\dagger} c_{j \sigma} 
                  + c_{j \sigma}^{\dagger} c_{j+1 \sigma}) + U \sum_{j=1}^L   
    (n_{j \uparrow}-\frac{1}{2})(n_{j \downarrow}-\frac{1}{2}),
\label{hamiltonian.hubbard}
\end{equation}
has been intensively studied in condensed matter physics as one of the
fundamental models for strongly correlated electrons. In 1968, Lieb
and Wu \cite{Lieb68} solved the model exactly by means of the 
coordinate Bethe ansatz method. Based on the Bethe ansatz, many 
physical quantities have been studied 
\cite{Korepin93,Takahashi99,Deguchi00}.  Especially in 1972, Takahashi 
introduced the string hypothesis and derived the thermodynamic Bethe 
ansatz equations (TBA) \cite{Takahashi72}, from which several thermal 
properties were calculated for finite temperature 
\cite{Takahashi74,Kawakami89,Usuki90,Takahashi02}. 

More recently quite a different approach to the thermodynamics of the 1D
Hubbard model has been developed based on the quantum transfer matrix (QTM)
method \cite{Suzuki87,Koma87,Inoue88,Wadati88,Koma8990,Yamada90,JSuzuki90,Takahashi91,Tsunetsugu91,JSuzuki92,Kluemper9293,Mizuta94,Kluemper96,Mizuta97,Juettner98,Juettner98_2,Kuniba98,Sakai99,Sakai99_2,Kluemper01,Sakai01}. In this approach, the problem is reduced to the
eigenvalue problem of the QTM. The free energy and other bulk quantities can
be calculated from the largest eigenvalue of the QTM. In 1998, J\"{u}ttner,
Kl\"{u}mper and Suzuki constructed the QTM utilizing fully the integrability
structure of the Hubbard model \cite{Juettner98_2}. For the largest eigenvalue
they derived a new type of non-linear integral equations (NLIE).  This set of
NLIEs turned out to be very powerful especially for the numerical study of the
thermal quantities, since it consists of only three auxiliary functions.

One of the remarkable advantages of the QTM approach is that it 
enables us to study the asymptotics of several correlation functions 
at finite temperature.  
Explicit results for the correlation lengths can be 
derived from the ratios of the next-leading eigenvalues to the largest 
eigenvalue of the QTM. Using this idea, Tsunetsugu calculated the correlation 
length of the spin-spin 
correlations for the half-filled 1D Hubbard model by 
solving the Bethe Ansatz equations numerically \cite{Tsunetsugu91}. In
the low temperature limit, his results agree with the finite 
temperature correction of the conformal field theory as $ \xi_{\rm s}
\propto v_{\rm s}/T $.  Later this was confirmed analytically by
Kl\"{u}mper and Bariev \cite{Kluemper96}.

In this paper, we study another important correlation function of the
half-filled 1D Hubbard model, namely the correlation length of the equal-time
one-particle Green's function ${\langle c_{k \sigma}^{\dagger} c_{j \sigma}
  \rangle}$. As is well known, the system is an insulator at half-filling and
the charge excitation has a gap. There is no effect of the charge gap on the
spin-spin correlation function, because the charge excitation is not relevant
for the spin-spin correlations.  Quite differently, the charge gap plays a
significant role for the one-particle Green's function leading to exponential
asymptotics even at ${T=0}$.  The quantitative relation, however, between the
charge gap and the correlation length is absolutely non-trivial and not well
understood yet.  In regard to this point, Stafford and Millis
\cite{Stafford93} conjectured that the correlation length ${\xi}$ for the
one-particle Green's function at ${T=0}$ is given by
\begin{equation}
1/\xi = \frac{4}{U} \int_{1}^{\infty} 
\frac{\ln(y + \sqrt{y^2-1})}{\cosh(2\pi y/U)} {\rm d} y.  \label{stafford}
\end{equation}
They obtained formula (\ref{stafford}) in a rather heuristic way from
an analytic calculation of the finite size ${L}$ dependence of the Drude
weight at ${T=0}$ \cite{Stafford93}
\begin{equation}
D_{\rm c}(L)=(-1)^{L/2+1}L^{1/2}D(U)e^{-L/\xi_{\rm c}(U)}.
\end{equation}

Stafford and Millis argued that the correlation length of the
one-particle Green's function is identical to the length
${\xi_{\rm c}(U)}$ whose computation yielded expression (\ref{stafford}).  
Certainly, in the limiting
cases ${U \rightarrow 0}$ and ${U \rightarrow \infty}$ 
formula (\ref{stafford}) shows the expected behaviour
\cite{Stafford93}. However, the status of (\ref{stafford}) is 
that of a conjecture, especially for finite values of ${U}$, as a proof
is still missing for the identity of the finite size scaling length
$\xi_{\rm c}$ and the Green's function correlation length ${\xi}$.
One of the goals of this paper is to confirm formula
(\ref{stafford}) directly by means of the QTM method and numerical
calculations.

In ref.~\cite{Juettner98_2}, J\"{u}ttner {\it et al} 
utilized Shastry's ${R}$-matrix
\cite{Shastry86,Olmedilla87,Shastry88,Shiroishi95} to construct the 
QTM for the 1D Hubbard model. Therefore, their Bethe ansatz equations 
are strictly valid only for the coupled spin model,
which is obtained from the 1D Hubbard model through the Jordan-Wigner 
transformation.  
This transformation poses no problem for the
investigation of the bulk properties of the system or the asymptotics of 
correlation functions such as the spin-spin correlations. 
However, for studying genuinely fermionic correlation functions, especially,
the one-particle Green's function, it is necessary to take properly 
into account its fermionic nature. For this purpose, we employ here 
the fermionic formulation of the QTM developed by one of the authors 
\cite{Umeno01,Umeno02}. 
(A related fermionic QTM for the lattice
spinless fermion model was studied in ref.~\cite{Sakai99}.) 
Below we sketch the salient points of the fermionic QTM for the 1D Hubbard model.
First we introduce the fermionic ${R}$-operator for the 1D Hubbard
model \cite{Umeno01,Umeno02,Umeno98} defined by
\begin{eqnarray}
\fR_{jk}(u,v)&=& \fR^{(\uparrow)}_{jk}(u-v) \fR^{(\downarrow)}_{jk}(u-v) 
+  \frac{\cos (u-v)}{\cos (u+v)} \tanh (h(u) - h(v)) \fR^{(\uparrow)}_{jk}(u+v) 
\nonumber \\
& & \hspace{4cm} \times \fR^{(\downarrow)}_{jk}(u+v)
(2n_{j \uparrow}-1)(2n_{j \downarrow}-1),
\label{fermionicRhubbard} \\
\fR^{(\sigma)}_{jk}(u)&=&
 \ a(u)(-n_{j \sigma}n_{k \sigma}+(1-n_{j \sigma})(1-n_{k \sigma})) 
-b(u)(n_{j \sigma}(1-n_{k \sigma})+(1-n_{j \sigma})n_{k \sigma})
\nonumber\\
&&+c(u)(c^{\dagger}_{j \sigma}c_{k \sigma}+c^{\dagger}_{k \sigma}c_{j \sigma}
), \ \ \ \ (\sigma=\uparrow, \downarrow)
\end{eqnarray}
where
\begin{equation}
a(u)=\cos u, \ \ b(u)=\sin u, \ \  c(u)=1, \ \  
 \frac{\sinh 2h(u)}{\sin 2u}=\frac{U}{4}.
\end{equation}
Due to the Yang-Baxter equation for the fermionic ${R}$-operator \cite{Umeno98}
\begin{equation}
\fR_{12}(u_1,u_2) \fR_{13}(u_1,u_3) \fR_{23}(u_2,u_3) \nonumber \\
= \fR_{23}(u_2,u_3) \fR_{13}(u_1,u_3) \fR_{12}(u_1,u_2),
\label{ybe}
\end{equation}
 the row-to-row transfer matrix, ${
\tau(u) = {\rm Str}_a \left\{ \fR_{aL}(u,0) \cdots \fR_{a1}(u,0) \right\} \label{monodromy2}
}$
constitutes a commuting family of operators, ${[\tau(u),\tau(v)]=0}$. 
Hence from the logarithmic derivative of the transfer matrix, we obtain the
set of commuting local fermionic operators ${\{I^{(n)}\}}$,
\begin{eqnarray}
  &\tau(u) = \tau(0) \exp \Big\{ u I^{(1)}+ \frac{u^2}{2!} I^{(2)}+
  \frac{u^3}{3!} I^{(3)} + \cdots
  \Big\},\label{conservedcurrent}
\end{eqnarray}
where $I^{(1)}$ is nothing but the Hamiltonian ${\cal{H}}$ (\ref{hamiltonian.hubbard}).
Next, following ref.~\cite{Umeno01} we introduce the super-transposed ${R}$-operator
${\bar{\fR}_{aj}(u_j,u_a) \equiv \fR_{ja}^{{\rm st}_a}(u_j,u_a)}$, as well as
a conjugated transfer matrix ${
\bar{\tau}(u) = {\rm Str}_a \left\{ \bar{\fR}_{aL}(0,u) \cdots \bar{\fR}_{a1}(0,u)
\right\}. \label{transfermatrix2}
}$
(See the original reference \cite{Umeno01} for the definition of
``${\rm st}$'' and the explicit form of ${\bar{\fR}_{aj}(u_j,u_a)}$.)
Since ${\bar{\tau}(u)}$ is actually expressed as
\begin{eqnarray}
\bar{\tau}(u)
        &=& {\rm Str}_a  \left\{ \fR_{1a}(0,u) \cdots \fR_{La}(0,u) \right\}, 
\end{eqnarray}
we can easily deduce 
\begin{equation}
  \bar{\tau}(u)= \tau(0)^{-1} \exp \left\{-u \mathcal{H} + O(u^2) \right\}. \label{conservedcurrent2}
\end{equation}

Combining (\ref{conservedcurrent}) and (\ref{conservedcurrent2}) we obtain the
fundamental relation
\begin{equation}
\tau(-u) \bar{\tau} (u) = \exp \left\{ - 2 u \mathcal{H} + O(u^2) \right\},
\end{equation}
which allows us to represent the partition function of the system as
\begin{eqnarray}
  Z &=& {\rm Tr} \ {\rm e}^{- \beta \mathcal{H}} = \lim_{N \rightarrow
  \infty} {\rm Tr} \left\{ \tau(-u_N) \bar{\tau} (u_N) \right\}^{N/2}
  \nonumber \\ 
    &=& \lim_{N \rightarrow \infty} {\rm Str} \left\{
  \tau^{\rm QTM}(u_N,0) \right\}^L, \hspace{1cm} \left( u_N =
  \frac{\beta}{N} \right).
\end{eqnarray}
Here ${\beta}$ is the inverse temperature and ${N}$ is called the
Trotter number.  The quantum transfer matrix ${\tau^{\rm QTM}(u_N,v)}$
is defined by
\begin{equation}
\tau^{\rm QTM}(u_N,v)={\rm Tr}_j \Big\{ \fR_{a_{N}j}(-u_N,v) \bar{\fR}_{a_{N-1}j}(v,u_N)
\cdots \fR_{a_{2}j}(-u_N,v) \bar{\fR}_{a_{1}j}(v,u_N) \Big\}. 
\label{qtm}
\end{equation}

From the largest eigenvalue
${\Lambda_0}$ of ${\tau^{\rm QTM}(u_N,0)}$, the free energy per site is
obtained in the thermodynamic limit as
\begin{equation}
\beta f = - \lim_{N \rightarrow \infty} \ln \Lambda_0.
\end{equation}
The correlation lengths ${\xi_i}$ of 
the (static) correlation functions can be evaluated 
from the next-leading eigenvalues 
\begin{equation}
\frac{1}{\xi_i} = - \Re \ln (\Lambda_i/\Lambda_0).
\end{equation}
Though the largest eigenvalue ${\Lambda_0}$ is always real and
positive, the next leading eigenvalues are not necessarily so. 
In this case, $ \Im \ln (\Lambda_i/\Lambda_0)$ is responsible
for oscillations of the corresponding correlation function.

The fermionic QTM (\ref{qtm}) has been diagonalized \cite{Umeno01} by
means of the quantum inverse scattering method based on the technique
by Ramos and Martins \cite{Ramos97,Martins98}.  The eigenstates of the
QTM are described by the set of rapidities ${\{s_{j},w_{l}\},
(j=1,\cdots,n, l=1,\cdots, m)}$, satisfying the Bethe ansatz equations
\begin{equation}
 \phi(s_j) = 
-\frac{q_2(s_j-i\gamma)}{q_2(s_j+i\gamma)}, \ \ \ \ 
\frac{q_2(w_l+2i\gamma)}
{q_2(w_l-2i\gamma)} = -
\frac{q_1(w_l+i\gamma)}{q_1(w_l-i\gamma)},
\label{bae}
\end{equation}

\begin{eqnarray}
q_1(s) &=& \prod_j^n (s-s_j), \quad q_2(s) = \prod_l^m (s-w_l), \nonumber \\
\phi(s)&=&
\left( - \frac{(1-z_{-}/z(s))(1-z_{+}/z(s))}
{(1+z_{-}/z(s))(1+z_{+}/z(s))} \right)^{N/2},
\label{qhub}
\end{eqnarray}
with the following settings
\begin{equation}
\gamma =  \frac{U}{4}, \ \ z_{\pm} =  \exp(\alpha) \  \{ \tan u_N \}^{\pm 1},
\end{equation} 
and 
\begin{equation} 
z(s) = i s(1+\sqrt{(1-1/s^2)}), \ \ 
\sinh(\alpha) = - \frac{U}{4} \sin(2 u_N).
\end{equation}
Here, for convenience, we have applied the partial particle-hole
transformation to the Bethe ansatz equations derived in
\cite{Umeno01}. Note that the one-particle Green's function is
``invariant" under the partial particle-hole transformation.

The corresponding eigenvalue of the fermionic QTM, ${\tau^{\rm
QTM}(u_N,0)}$ is given by
\begin{equation}
\Lambda = c(u_N) \prod_{j=1}^{n} z_j, \label{eigenvalue}
\end{equation}
where ${z_j \equiv z(s_j)}$. We omit here the precise form of the
prefactor ${c(u_N)}$, which is in common for all eigenstates and is not
necessary for the evaluation of the correlation lengths.

\begin{figure}[htbp]
\twoimages[scale=0.7]{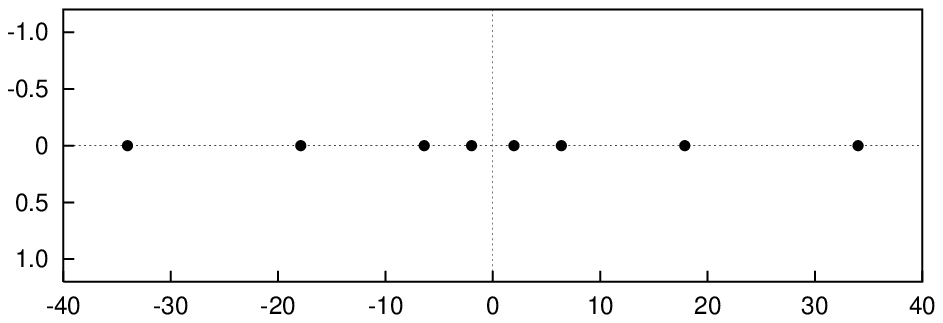} {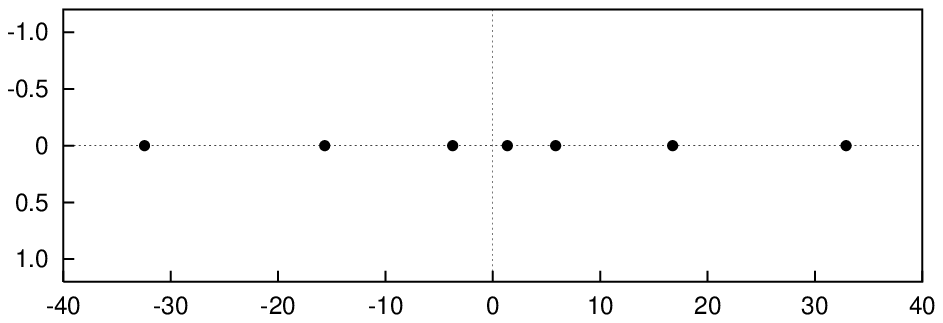}
\caption{(a) The rapidities ${z_j}$ on the imaginary axis
  (horizontal line) for the largest eigenvalue ${\Lambda_0}$ in the sector
  ${N=8}$, ${n=8}$,${m=4}$, ${U=8, \beta = 0.8}$. (b) 
 The rapidities ${z_j}$ on the imaginary axis
  (horizontal line) for the largest eigenvalue ${\Lambda_{\rm G}}$ in the
  sector ${N=8}$, ${n=7}$, ${m=3}$, ${U=8, \beta = 0.8}$}
\label{fig:1}
\end{figure}

We like to emphasize that the 
fermionic statistics is properly incorporated in
the Bethe ansatz equations (\ref{bae}) and the eigenvalue formula
(\ref{eigenvalue}). For example, we have found that the dominant eigenvalue 
contributing to the correlation length for ${\langle c^{\dagger}_{k\sigma}
  c_{j\sigma} \rangle}$ is the largest eigenvalue in the sector $[n=N-1,
m=N/2-1]$. In Fig.~\ref{fig:1} (b), we plot an example of the distribution pattern
of the rapidities ${z_j}$ for the dominant eigenvalue ${\Lambda_{\rm G}}$. As
all rapidities ${z_j}$ turn out to be 
purely imaginary, we plot them in a
90-degree rotated frame, i.e., the horizontal line is the imaginary axis.
Compared with ${z_j}$ for the largest eigenvalue $\Lambda_0$ (see Fig.
\ref{fig:1} (a)), we can roughly say that one of the ${z_j}$ closest to the origin is removed.  Note that the complex conjugate of ${\Lambda_{\rm G}}$ is also
the largest eigenvalue of the sector $[n=N-1, m=N/2-1]$ and the corresponding
rapidities ${z_j}$ are given by the reflection of those for ${\Lambda_{\rm
    G}}$ with respect to the real axis (the vertical line in Fig.~\ref{fig:1}).
From the expression (\ref{eigenvalue}) and the distribution pattern of ${z_j}$
in Fig.~\ref{fig:1} (b), we find $\Lambda_{\rm G}$ to be purely imaginary. This is consistent with the fact that the one-particle Green's function should exhibit ${k_{\rm F}=\pi/2}$ oscillations at half-filling.
\begin{figure}[htbp]
\twoimages[width=0.49\textwidth,height=0.22\textheight]{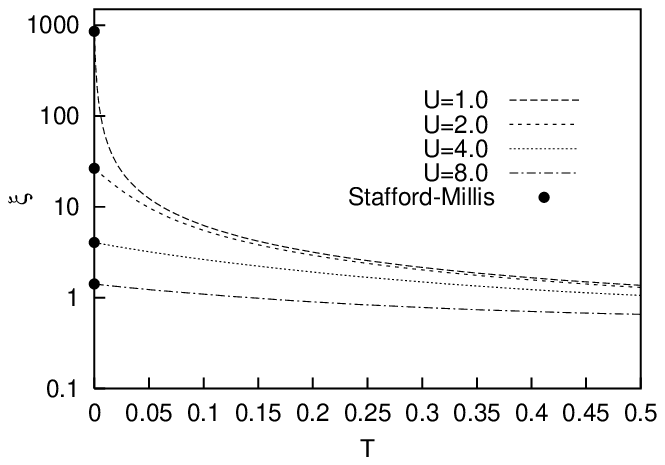}{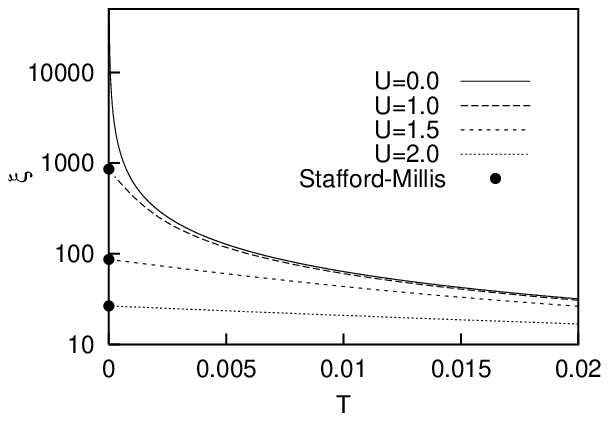}
\caption{(a) Temperature dependence of the correlation
lengths for one-particle Green's function for the half-filled Hubbard
model. (b) Similar to (a) at lower temperatures}
\label{fig:3}
\end{figure}
We have numerically solved the Bethe ansatz equations (\ref{bae}) for
increasing Trotter numbers ${N}$ and calculated the correlation length
${\xi}$ from the ratios of the eigenvalues ${\Lambda_{\rm G}}$ and
${\Lambda_0}$.  The result is plotted in Fig.~\ref{fig:3}. 
The largest Trotter number we took is ${N=10240}$ down to temperatures as low
as ${T=0.001}$. For such low temperatures, we have extrapolated our data for
${N=6144,8192,10240}$ to estimate the limiting behaviour for ${N \rightarrow
  \infty}$. The error of the extrapolated results is less than 0.5 \%. For
moderate temperatures (${T > 0.01}$), it was sufficient to use Trotter numbers
just about ${N=4096}$ to get very accurate data.

\begin{figure}[htbp]
\twoimages[width=0.49\textwidth,height=0.22\textheight]{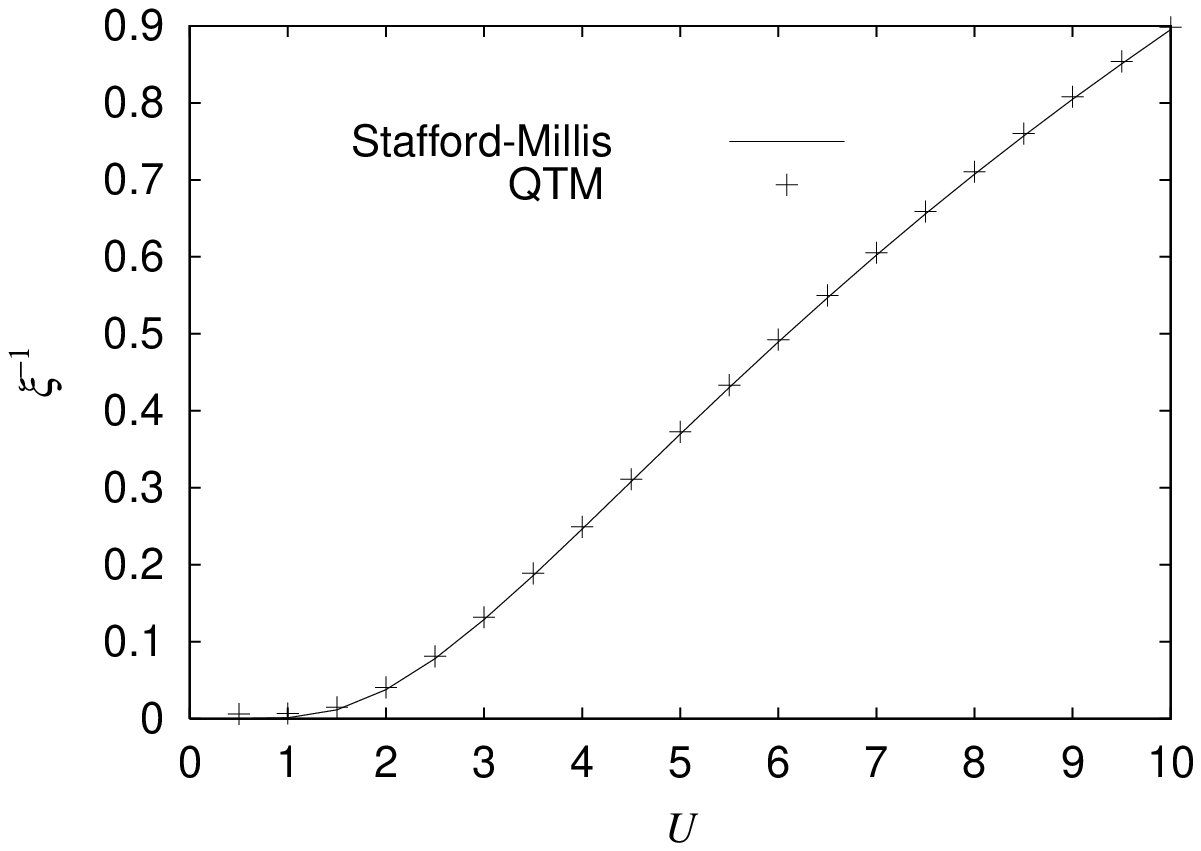}{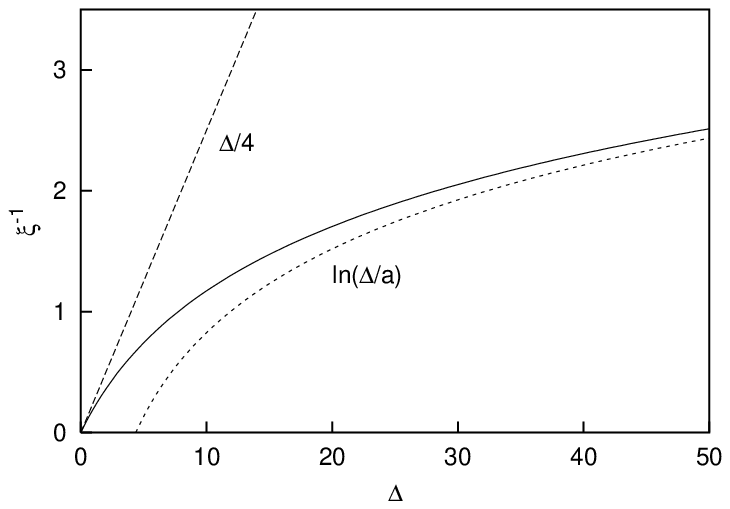}
\caption{(a) Comparison of data obtained from an extrapolation of the
  finite temperature correlation length to ${T=0}$ with Stafford and Millis'
  conjecture. (b) The dependence of the ${T=0}$ correlation length ${\xi}$ on the
  charge gap ${\Delta}$. The dotted lines correspond to the asymptotics
  (\ref{asym1}) and (\ref{asym2}).}
\label{fig:4}
\end{figure}

From Fig.~\ref{fig:3}, we can clearly observe that the correlation lengths
remain finite as ${T \rightarrow 0}$.  The extrapolation of our data to ${T
  \rightarrow 0}$ is compared with (\ref{stafford}) in Fig.~\ref{fig:4} (a).
To perform the extrapolation, we assumed the low temperature behaviour of the
correlation length in the form ${\xi = \xi_{T=0} \times \exp (- \gamma T)}$
and estimated ${\xi_{T=0}}$ from our numerical data at low temperatures.
Obviously our extrapolations coincide with formula (\ref{stafford}) almost
perfectly.  Therefore we conclude Stafford and Millis' conjecture is correct
for {\it any} value of ${U}$.

Having confirmed that formula (\ref{stafford}) gives the correct
correlation length ${\xi}$ at ${T=0}$, we now look at 
its dependence on the gap ${\Delta}$, see Fig.~\ref{fig:4} (b).  
Here, we used the well known gap formula
\begin{equation}
  \Delta = \frac{16}{U} \int_{1}^{\infty} \frac{\sqrt{y^2-1}}{\sinh (2 \pi
    y/U)} {\rm d} y. \label{gap}
\end{equation}
Usually the correlation length of a gapped system is believed to be
simply inversely proportional to the magnitude of the gap. 
Equations (\ref{stafford}) and (\ref{gap}) indeed give ${\xi^{-1}
\propto \Delta}$ for small ${U}$, however ${\xi^{-1} \propto \ln \Delta (\sim
\ln U) }$ for large ${U}$ \cite{Stafford93}. 
More precisely, we have 
\begin{eqnarray}
\xi^{-1}& & \sim \Delta/4 \ \ \ \ \ \ \ \ \ \ \ (U \rightarrow 0), \label{asym1} \\
\xi^{-1}& & \sim \ln (\Delta/a) \ \ \ \ \ \ (U \rightarrow \infty), \label{asym2} 
\end{eqnarray}
where ${a = [\Gamma(\frac{1}{4})/\sqrt{2 \pi}]^4 \simeq 4.37688}$ (see
ref.~\cite{Stafford93}).  Thus the gap dependence of the correlation length
does not follow a simple rule.  For a discussion of this issue we like to refer  the reader to ref.~\cite{Okunishi01}.

Finally, for increasing temperature ${T}$, the correlation
length ${\xi}$ is getting closer to the non-interacting free-fermion 
(${U}$=0) value, namely ${\xi_{U=0} = 1/ \ln \left(\pi T/2 + \sqrt{(\pi T/2)^2+1} 
\right)}$, irrespective of the value of ${U}$.
Actually, for small ${U (\sim 1)}$, the correlation
length ${\xi}$ takes almost the same value as ${\xi_{U=0}}$ 
even at rather low temperatures (${T \sim 0.01}$) 

In this paper we have presented a numerical treatment of the one-particle
correlation length $\xi(T)$ for finite Trotter number $N$.  In a future
publication, we will derive the NLIE for the corresponding eigenvalue
${\Lambda_{\rm G}}$ in the limit $N\to\infty$. This will allow us to obtain
analytic results for the $T\to 0$ limit and to prove formula (\ref{stafford})
analytically. 

After completing this manuscript we found a derivation of the
zero temperature correlation length (\ref{stafford}) on the basis of the
standard row-to-row transfer matrix of the Hubbard model and the
one-particle energy-momentum excitation following the method described in
\cite{Okunishi01}. Details of this calculation will be published elsewhere.

\acknowledgments
The authors are grateful to M. Wadati, M. Takahashi, M. Inoue, K. Sakai and K.
Saito for valuable discussions and continuous encouragements. They also thank
the Supercomputer Center, Institute for Solid State Physics, University of
Tokyo for the use of the facilities.

\end{document}